\documentclass[twocolumn,showpacs,preprintnumbers,amsmath,amssymb,prb,superscriptaddress]{revtex4} 
 
\usepackage{epsfig}
\usepackage{graphicx}
\usepackage{dcolumn}
\usepackage{bm}
 
\begin{document} 
 
\title{Static impurities in the kagome lattice: dimer freezing and mutual  
repulsion} 
 
\author{S. Dommange}
\affiliation{Institut Romand de Recherche Num\'erique en Physique des  
Materiaux (IRRMA), PPH--Ecublens, CH--1015 Lausanne, Switzerland}  
 
\author{M. Mambrini} 
\affiliation{Laboratoire de Physique Th\'eorique, Universit\'e Paul Sabatier,  
FRE 2603, F--31062 Toulouse Cedex, France} 
 
\author{B. Normand} 
\affiliation{D\'epartement de Physique, Universit\'e de Fribourg,  
CH--1700 Fribourg, Switzerland} 
 
\author{F. Mila} 
\affiliation{Institut de Physique Th\'eorique, Universit\'e de Lausanne,  
CH--1015 Lausanne, Switzerland} 
 
\date{\today} 
 
\begin{abstract} 
 
We consider the effects of doping the $S$ = 1/2 kagome lattice with  
static impurities. We demonstrate that impurities lower the number 
of low--lying singlet states, induce dimer--dimer correlations  
of considerable spatial extent, and do not generate free spin degrees  
of freedom. Most importantly, they experience a highly unconventional  
mutual repulsion as a direct consequence of the strong spin frustration.  
These properties are illustrated by exact diagonalization, and reproduced 
to semi--quantitative accuracy within a dimer resonating--valence--bond 
description which affords access to longer length scales. We calculate 
the local magnetization induced by doped impurities, and consider its 
implications for nuclear magnetic resonance measurements on known kagome 
systems. 
 
\end{abstract} 
 
 
\maketitle 
 
\section{Introduction} 
 
The kagome geometry (Fig.~1) presents one of the most highly frustrated 
quantum spin systems achievable in two dimensions (2d) with only 
nearest--neighbor Heisenberg interactions. The $S$ = 1/2 kagome lattice 
has been found to have a spin--liquid ground state,\cite{rce} whose 
ultra--short spin--spin correlation lengths are effectively those of a dimer 
liquid,\cite{rle,rze} and whose excitation spectrum\cite{rlblps} shows a 
manifold containing an extremely large number of low--lying singlets.\cite{rm} 
These properties are reproduced very well by a short--range 
resonating--valence--bond (RVB) description,\cite{rmm} based in fact 
only on nearest--neighbor dimer formation (Fig.~1).

On the theoretical level, a similar degree of frustration is found only 
in quantum dimer models,\cite{rrk,rms} which have also been considered 
recently in the kagome geometry,\cite{rmsp} and in models containing 
higher--order spin terms and multiple--spin exchange.\cite{rmblw}
Experimentally, a variety of materials displaying the kagome structure 
is known to exist, although to date none have been found which contain
an ideal, 2d, vacancy--free system of spins $S = 1/2$. These 
include the jarosites (H$_3$O)Fe$_3$(OH)$_6$(SO$_4$)$_2$,\cite{rwhrs} 
KFe$_3$(OH)$_6$(SO$_4$)$_2$, and KCr$_3$(OH)$_6$(SO$_4$)$_2$\cite{rkea,rlea}
and the volborthite Cu$_3$V$_2$O$_7$(OH)$_2$.H$_2$O.\cite{rhhkntkt}
The best characterized of these compounds is the magnetoplombite 
SrCr$_{9p}$Ga$_{12-9p}$O$_{19}$ (SCGO),\cite{rlmcmombm} in which some of 
the $S$ = 3/2 Cr$^{3+}$ ions form approximately 2d kagome planes. However, 
because in this case the ideal stoichiometry ($p = 0$) remains unachievable, 
discussion of the influence of static (and also, for Ga$^{3+}$, spinless) 
impurities represents an important issue.

\begin{figure}[t!] 
\medskip
\centerline{\psfig{figure=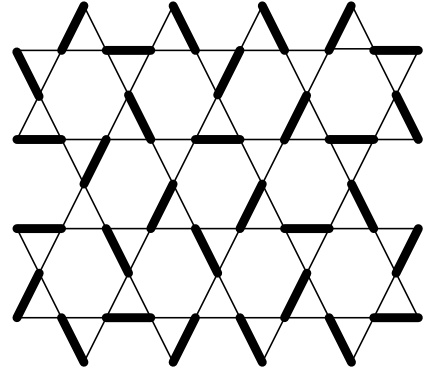,width=5.5cm,angle=0}}
\medskip
\caption{The kagome lattice. Thick black lines represent an arbitrarily 
chosen covering by nearest--neighbor dimers formed between $S = 1/2$ spins 
on each lattice site. } 
\label{fig:1} 
\end{figure} 
 
Studies of static vacancies in spin liquids have tended to focus on 
the question of induced free spin degrees of freedom.\cite{rlmgmdhlr} 
In systems such as spin ladders, and also for the 2d square 
lattice,\cite{rbhsl} the total energy appears to decrease as such holes 
in the spin background are made to approach each other, implying an 
effective mutual attraction. It is generally believed that in an RVB 
description of the ground state the minimal disruption of the wave 
function, and thus the most favorable energy state, occurs when holes 
occupy adjacent sites, thus generating a simple pairing mechanism for 
mobile dopants. The issue of impurites in frustrated magnets has received 
less attention, and appears to be more involved: it has been proposed that 
an absence of induced free spins corresponds to deconfined, spinon-like 
excitations.\cite{rsv} Consideration of the problem for the 1d frustrated 
chain,\cite{rnm} which provides an example with no free spins localized 
around the vacancies, suggests the importance of degenerate singlets in 
the ground--state manifold. This result indicates that the kagome geometry 
presents the most probable candidate for exotic behavior\cite{rs} as a
consequence of strong frustration in 2d.  
 
In this study we investigate the effects of static impurities in the kagome 
lattice. We find that, despite the very short intrinsic spin--spin 
correlation lengths, dimer--dimer correlations develop over considerable 
distances in the presence of dopants. Further, we show that there is no 
evidence for free local moments induced around impurity sites, and that 
a highly unusual effective repulsive interaction arises between such 
holes in the spin background. 

In Sec.~II we provide an outline of the numerical and anaytical methods 
to be employed, and discuss the fundamental effects on the excitation 
spectrum of doping by spinless impurities. In Sec.~III we consider the 
case of one hole in an odd--site cluster, demonstrating ``dimer freezing'' 
in the spin--spin correlation functions and dimer--dimer correlations of 
anomalously long range. The induced magnetizations in the spin sectors 
$S$ = 1 and higher are used to illustrate the consequences for nuclear 
magnetic resonance (NMR) measurements. Sec.~IV presents spin--spin 
correlation functions which characterize the situation with two impurities, 
supplemented by total--energy analyses quantifying the novel repulsive 
interaction between the holes which emerges from this study. Sec.~V 
contains further discussion and our conclusions.  
 
\section{Elimination of singlet states and absence of localized spins}
 
\subsection{Exact Diagonalization}

By ``impurity'' we refer henceforth to nonmagnetic impurities. The 
techniques we employ are based primarily on numerical calculations for 
small clusters with periodic boundary conditions chosen as symmetrically 
as possible. Specifically, we perform exact diagonalization (ED) studies 
of the Heisenberg Hamiltonian 
\begin{equation}
H = J \sum_{\langle ij \rangle} {\bf S}_i {\bf \cdot S}_j ,
\label{ehh}
\end{equation}
where $J$ is the antiferromagnetic superexchange interaction and $\langle 
ij \rangle$ denotes nearest--neighbor sites, for clusters up to 27 sites 
with one impurity, and up to 24 sites with two impurities for all available 
configurations. We analyze the total energies, the energies per doped 
impurity, and also the form of the energy spectra, focusing on the number 
of singlet states in the low--energy manifold below the first triplet for 
the finite systems under consideration, as well as on the number of these 
states removed by impurity doping. The frustrated spin interactions in the 
presence of impurities are characterized for the entire cluster by the bond 
spin--spin correlation functions and by the induced site magnetizations 
in every spin sector. 

\begin{figure}[t!] 
\centerline{\psfig{figure=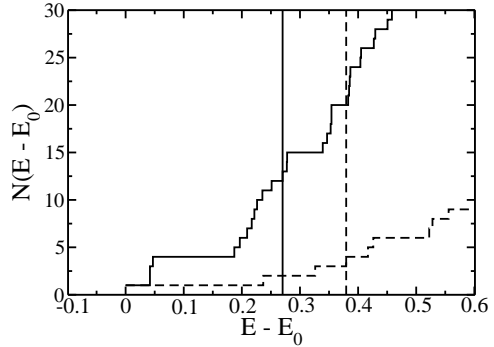,width=7.0cm,angle=0}}
\medskip
\caption{Integrated density of states for an 18--site $S$ = 1/2 kagome 
cluster, computed by exact diagonalization both without impurities (solid 
line) and with 2 impurities (dashed line). The vertical lines mark the 
position of the first triplet state in the absence of impurities (solid) 
and in their presence (dashed). } 
\label{fig:2} 
\end{figure} 

The density of states integrated over the lowest--lying energy levels 
is shown in Fig.~2 for systems with and without impurities. Fig.~2 
compares ED results from systems of 18 sites, for which in the impurity 
configuration illustrated 8 of the 12 singlet states occurring below 
the first triplet, indicated by the vertical lines, are eliminated by 
the presence of 2 impurities. Although the separation of the triplet 
from the lowest remaining singlet decreases, this energy gap always 
remains a finite fraction of the superexchange interaction $J$. Thus the 
most important qualitative result emerging even for the smallest systems is 
that, in contrast to spin--ladder and dimerized or quadrumerized systems, 
no low--lying triplet state is induced by the presence of impurities. This 
implies directly\cite{rnm} that there are no quasi--free spin degrees of 
freedom, by which is meant spins effectively isolated from their local 
environment and interacting with each other only on an effective energy 
scale much smaller than $J$. The fact that no localized moments are formed 
in the vicinity of impurities may be ascribed to the continuum of singlets 
beyond the ground state, which has the same function as the two--fold 
degeneracy of the ground state in the frustrated chain,\cite{rnm} namely 
that the rearrangement of dimers incurs no energy penalty. The absence of 
local spins around nonmagnetic impurities represents one of the most 
significant differences between frustrated and unfrustrated systems, and 
has been further interpreted as implying a deconfinement of the spin 
degrees of freedom.\cite{rsv}

\begin{table}[b!]
  \begin{tabular}{|c|cccc|ccccc|} \hline 
$N$ & \multicolumn{4}{c|}{12} & \multicolumn{5}{c|}{18} \\
\hline
$d_{\rm M}$ & 1 & 2 & 2 & 3 & 1 & 2 & 2 & 3 & 3 \\
$N_s^0$ & 4 & 4 & 4 & 4 & 12 & 12 & 12 & 12 & 12 \\
$N_s^2$ & 2 & 2 & 3 & 1 & 3 & 4 & 4 & 3 & 4 \\
\hline
\end{tabular}
\caption{\label{tab:deltaE} Number of low--lying singlets below the first 
triplet for kagome clusters without impurities ($N_s^0$, third row) and 
with two impurities ($N_s^2$, fourth row) as a function of the Manhattan 
distance $d_{\rm M}$ between impurities. Results obtained by ED are shown 
for clusters of $N$ = 12 and 18 sites. }
\end{table}

Table 1 provides more details of the number of states in the singlet 
manifold which are removed by impurity doping for clusters of 12 and 18 
sites, for all possible configurations of 2 impurities. The numbers vary 
little as a function of the proximity of the impurities, and the average 
number of low--lying states which are eliminated is in agreement with simple 
considerations of the number of singlet states in the small system (below). 
We have also performed analogous calculations for 24-site clusters of
different shapes. For all clusters the same behavior is found upon 
introducing two impurities, in that the number of singlets below the 
first triplet is reduced, but here the lowest--lying excitation is not 
a triplet. However, these clusters do not appear to be representative of 
the generic kagome physics, as the number of low--lying singlets in the 
absence of impurities is significantly smaller than the expected 
value\cite{rlblps} of $1.15^N$. This situation is presumed to be a 
consequence of the low symmetry of the 24--site clusters, some of which 
are essentially 1d systems, and as a result the densities of states are 
not presented here.

\begin{figure}[t!] 
\centerline{\psfig{figure=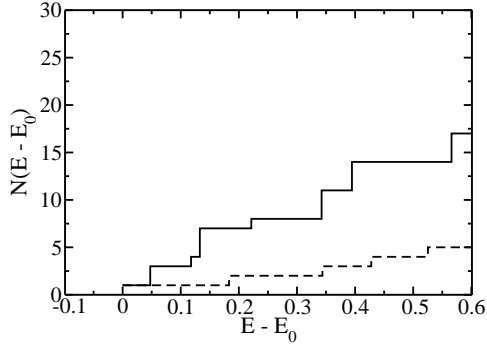,width=7.0cm,angle=0}}
\medskip
\caption{Integrated density of singlet states for an 18--site $S$ = 1/2 
kagome cluster, calculated in the dimer--RVB basis both without impurities 
(solid line) and with 2 impurities (dashed). } 
\label{fig:3} 
\end{figure} 
 
\subsection{Dimer--RVB Basis}

We calculate in addition the ground--state properties, namely the energy 
and spin--spin correlation functions, of clusters doped with one and with 
two impurities within the dimer--RVB basis. This is the basis of states in 
which every spin forms a singlet dimer with one of its nearest neighbors 
in such a way that every spin of the lattice is involved in a dimer 
(Fig.~1). The use of this basis\cite{rmm} naturally implies a truncation 
of the total number of available states in the manifold of low--lying 
singlets, but we will show below that it provides a satisfactory description 
of the ground state of the doped system. Dimer--RVB calculations permit 
the consideration of somewhat larger cluster sizes, and we will present 
results up to and including 39 sites with one impurity and 36 sites with 
two impurities.

Calculations performed in the dimer--RVB basis yield results very similar 
to the exact ones for the integrated density of states (Fig.~3), giving 
clear evidence of a strong linear correlation (see also Fig.~9) between 
the ED and dimer--RVB spectra in both 0-- and 2--impurity cases. 
Demonstration of a quantitative correspondence between the spectra 
would require shift and scaling factors arising from an energetic 
renormalization related to the fact that calculations in the dimer--RVB 
basis are variational in nature, so that the true eigenstates are dressed 
RVB wave functions, but this procedure is not necessary for the analysis 
to be presented here. 

Given that the RVB basis represents a very significant truncation of the 
total Hilbert space, the qualitative similarity of the dimer--RVB spectra 
to those obtained from the exact calculations, both with and without 
impurities, provides important information about the nature of the exact 
wave functions of the kagome system. This observation is further confirmed 
by the semi--quantitative agreement with exact results for spin correlation 
functions and ground--state energies computed using the dimer--RVB basis. 
We defer this comparison to Secs.~III and IV, and comment here only that 
our results in the presence of impurities provide additional strong 
justification for the statement that the ground state of the kagome spin 
system is described appropriately by a maximally short--ranged RVB wave 
function, presumably as a consequence of the very strong frustration. 

The dimer--RVB basis is a natural framework for understanding the absence 
of localized spins around an impurity site, as all remaining spins are
simply incorporated within rearranged singlet dimer coverings. It also 
provides a qualitative reflection of the ``evaporation'' of low--lying 
singlets from the ground--state manifold (Figs.~2--3 and Table I) which 
arises on doping a cluster with two impurities, and in fact yields 
additional insight into the mechanism for this exclusion. It can be 
shown that the total number of singlet dimer coverings of a doped system 
is given by 
\begin{equation}
N_{\rm coverings} = 2^{N/3 - N_{\rm imp} + N_G + N_{\rm et}},
\label{endc}
\end{equation}
where $N$ is the number of sites of the system under consideration, 
$N_{\rm imp}$ the number of doped impurities, $N_G$ the number of independent 
clusters into which the system is divided by the dopant distribution, and 
$N_{\rm et}$ the number of triangles containing 3 impurities. A complete 
derivation of this result lies beyond the scope of the current analysis, 
and we present only the following heuristic explanation. By reducing each 
triangle to an effective $S = 1/2$ degree of freedom,\cite{rm,rmm} the 
dimer basis involves a truncation of the $2^N$ total states to $2^{N/3}$, 
multiplied by an additional factor of 2 for the overall spin direction on 
each independent cluster. Each added impurity removes one spin, halving 
the number of states in every manifold and thus justifying the dependence 
on $N_{\rm imp}$. The only exception to this statement occurs when the 
added impurity falls on a triangle which already contains two impurities, 
as a result of which it is already fully constrained and removal of the 
third site no longer affects the number of coverings. Eq.~(\ref{endc}) 
can be shown to remain valid for all vacancy concentrations, including 
most notably values of $N_{\rm imp}$ exceeding the percolation threshold, 
beyond which $N_G > 1$, under the condition that the impurity distribution 
does not cut the system into segments containing odd numbers of spins.

\section{Spin correlation functions and induced magnetization}

\subsection{Dimer ``freezing''}

In this section we consider the effects of a single impurity introduced  
in an odd--site cluster. The bond spin--spin correlation functions 
$\langle {\bf S}_i {\bf \cdot S}_j \rangle$ around a single impurity 
may be computed by ED for clusters of up to 27 sites. The ED calculations 
permit in addition the determination of the magnetizations induced on each 
site by the presence of the impurity, with results for every spin sector 
corresponding to the situation in all applied magnetic fields, which 
are presented in Subsec.~C below. 

\begin{figure}[t!] 
\centerline{\psfig{figure=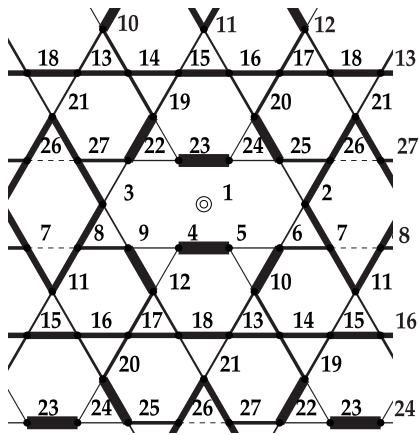,width=5.5cm,angle=0}}
\medskip
\caption{Spin--spin correlations for a single impurity in a 27--site  
cluster, obtained by ED. Bar width represents the strength of correlation 
functions on each bond on a linear scale where the strongest correlation 
function is $\langle {\bf S}_i {\bf \cdot S}_j \rangle = - 0.69$. Dashed 
lines denote bonds on which $\langle {\bf S}_i {\bf \cdot S}_j \rangle > 0$.}
\label{fig:4} 
\end{figure} 

The correlation functions obtained by ED are illustrated in Fig.~4 
for a 27--site cluster with periodic boundary conditions. The most 
striking feature of these results is that dimer bonds are effectively 
frozen in the two depleted triangles, taking a value $\langle {\bf S}_i 
\cdot {\bf  S}_j \rangle \simeq - 0.69$ close to the theoretical minimum 
of $-3/4$ for a pure singlet state. For comparison, the average value of 
$\langle {\bf S}_i \cdot {\bf  S}_j \rangle$ in the pure system, which 
retains full freedom to resonate, is a factor of 3 smaller 
($- 0.22$).\cite{rle} The bonds neighboring the frozen dimers then show 
very weak correlations, while the next neighbors are again enhanced, and 
the spin correlations oscillate with distance away from the impurity 
[also illustrated in Fig.~5(b)]. Thus although spin--spin correlations 
in the kagome system are extremely short--ranged,\cite{rle} induced 
dimer--dimer correlations show a considerable spatial extent, reaching 
in fact well beyond the sizes obtainable by studying small clusters using 
either ED or dimer--RVB techniques. For the limited data available it is 
unfortunately not possible to specify the functional form of the decay to 
equilibrium. Further analysis of this issue, based on data obtained from 
mutual hole repulsion, is presented in Sec.~IV. 

\begin{figure}[t!] 
\psfig{figure=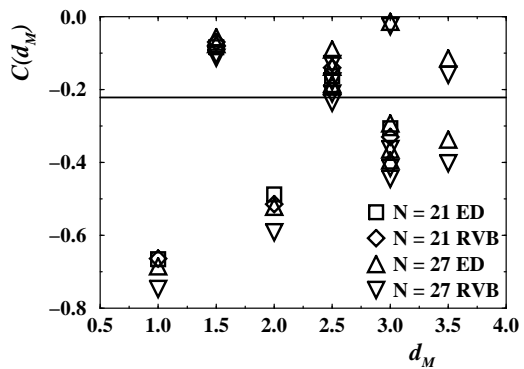,width=5.0cm,angle=270}
\centerline{(a)}
\centerline{\psfig{figure=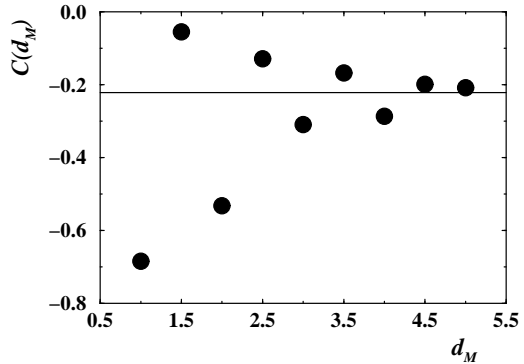,width=5.0cm,angle=270}}
\centerline{(b)}
\caption{(a) Spin--spin correlation functions $C(d_{\rm M}) = \langle 
{\bf S}_i \cdot {\bf  S}_j \rangle$ on nearest--neighbor bonds as a function 
of their Manhattan distance $d_{\rm M}$ from a single impurity, for 
clusters of 21 and 27 sites. (b) Spin--spin correlations as a function of 
(Manhattan) bond distance from a single impurity, obtained by dimer--RVB 
calculations for a 39--site cluster. Points represent the average over all 
bonds at the same distance from the impurity, the solid line the value 
$\langle {\bf S}_i {\bf \cdot S}_j \rangle = - 0.22$ for the impurity--free 
system.}
\label{fig:5} 
\end{figure} 
 
\subsection{Dimer--RVB Basis}

Fig.~5(a) compares the bond spin correlation values $\langle {\bf S}_i 
\cdot {\bf  S}_j \rangle$ obtained from ED and from the dimer--RVB basis 
for all bonds in clusters of 21 and 27 sites with one impurity. These 
clusters represent the maximal sizes accessible by ED, and for such 
systems it is clear that the dimer--RVB basis reproduces the exact 
values with quantitative accuracy. From the nature of the RVB basis it is 
perhaps to be expected that nearest--neighbor bond correlations are treated 
more accurately than any other physical property. We make use of this 
feature to extend the calculations of spin correlation functions to a 
system of 39 sites with oie impurity. Because on this cluster the boundary 
conditions lead to a rather low symmetry, Fig.~5(b) shows the value of 
$\langle {\bf S}_i {\bf \cdot S}_j \rangle$ averaged over all bonds at 
the same Manhattan distance from the impurity, which illustrates both 
the oscillatory nature of the frozen dimerization pattern and the very 
long range over which an impurity acts to disrupt the spin configuration.

\begin{figure}[t!] 
\centerline{\psfig{figure=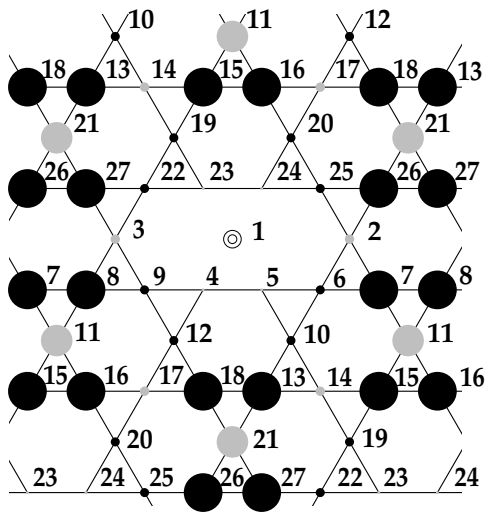,width=6.5cm,angle=0}}
\medskip
\caption{Site magnetization profile of the lowest triplet state induced 
by a single impurity in a 27--site cluster, obtained by ED. Dot radius 
represents the magnitude of the local moment on a linear scale where the 
largest dot corresponds to a moment of 0.13$\mu_{\rm B}$, and the grey dots 
represent sites with induced moments opposite to the field.} 
\label{fig:6} 
\end{figure} 
 
\subsection{Induced magnetic moments}

The local magnetization pattern induced in the vicinity of a doped  
nonmagnetic impurity may be computed by exact diagonalization in the sector 
of total spin $(S,S_z)$ = (1,1), and is displayed for the 27--site cluster 
in Fig.~6. The site magnetizations exhibit a certain anticorrelation with 
the strength of the dimers formed on each bond, and thus have a tendency 
to oscillate with increasing distance from the impurity. The induced 
moments are particularly enhanced on sites far from the impurity, where 
different paths on the periodic cluster interfere, and the interference 
in this regime may even result in moments oriented opposite to the 
effective magnetic field (Figs.~6, 7). The magnetization profile of the 
surrounding sites is the single most important consequence of impurities 
which is reflected in local--probe measurements of doped spin systems. 

\begin{figure}[t!] 
\psfig{figure=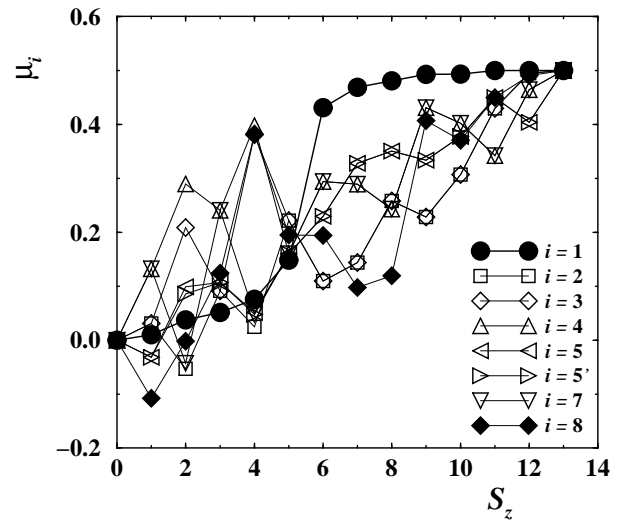,width=7.2cm,angle=270}
\medskip
\caption{Induced magnetization per site $\mu_i$ as a function of the spin 
sector $S_z$ (analogous to an applied magnetic field) for all inequivalent 
sites\cite{footnote1} neighboring an impurity, computed by ED for a 
27--site cluster. With reference to the site labels in Fig.~6, the 
separate lines $i$ represent 1st (4, 5, 23, 24), 2nd (6, 9, 22, 25), 3rd 
(10, 12, 19, 20), 4th (2, 3), equal 5th [(7, 8, 26, 27) and (13, 15, 16, 
18)], 7th (14, 17), and 8th (11, 21) neighbors of the impurity site (1).} 
\label{fig:7} 
\end{figure} 
  
Fig.~7 shows the magnetization per site induced by proximity to a 
nonmagnetic impurity for each spin sector obtainable by ED for a 
27--site cluster. The spin sector may be considered as equivalent to an 
effective magnetic field which varies between zero ($S_z = 0$) and 
saturation ($S_z = 13$). The most pronounced variation is observed for 
the nearest--neighbor sites (filled circles in Fig.~7), which have very 
small magnetization at low fields because they are bound into the strong 
dimers frozen by the presence of the impurity on the same triangle. 
However, between $S_z = 4$ and $S_z = 6$ the moment shows an abrupt jump 
to saturation, implying the breaking and full polarization of these dimers 
at fields in excess of 1/3 of the saturation field. We speculate that this 
behavior may be due to the presence of a particularly stable state of the 
remaining spins which corresponds to a 1/3 magnetization plateau of the 
undoped kagome lattice.\cite{rhcs} The moments of the 
further--neighbor sites show a generally less systematic evolution, 
usually with a non--monotonic character which implies a very significant 
rearrangement of the induced local magnetization as a function of field. 
This evolution is again marked by the most significant changes occurring 
close to the net magnetization value of 1/3. 

NMR has emerged as the technique of choice for analyzing in real space  
the effects of doped impurities on magnetizations and spin correlations. 
Information concerning these quantities is extracted from Knight--shifts, 
linewidth alterations and, for certain cases, from relaxation times. A  
detailed NMR and NQR (nuclear quadrupole resonance) analysis has been 
conducted recently for $^{71}$Ga impurities in the kagome planes of the $S$ 
= 3/2 spin system SrCr$_{9p}$Ga$_{12-9p}$O$_{19}$ (SCGO).\cite{rlmcmombm} 
While a quantitative determination of any shifts or linewidths depends  
on a detailed Hamiltonian for the hyperfine interactions in a specific  
material, the results obtained in Fig.~7 may be considered as illustrative  
for the kagome geometry. 

The important qualitative features are the absence of a single spin--1/2 
degree of freedom anywhere in the vicinity of the impurity, and the rather 
large extent of the area of affected sites in spite of the very short spin 
correlation lengths. The latter effect is contained in Fig.~7 in that the 
induced site magnetization values on the vertical axis are directly 
proportional to the peak shifts measured in an NMR field sweep. Impurities 
in the kagome lattice therefore produce a large number of satellite peaks. 

Finally, with regard to the fact that most known kagome systems, including 
SCGO, are not $S = 1/2$, there remain rather few studies of this geometry 
for higher values of the spin.\cite{rhcs,rrsr} On rather general grounds 
based on the geometry of the lattice, one expects that both triangles and 
depleted triangles retain the integral or half--integral nature of the 
spin per site. Thus it is at least plausible that the results for the 
$S$ = 1/2 system are directly relevant to higher half--integral--spin 
systems such as SCGO.\cite{footnote2} 

\begin{figure}[t!] 
\psfig{figure=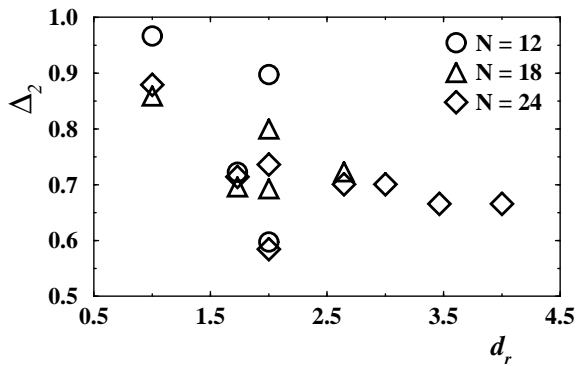,width=5.0cm,angle=270}
\medskip
\caption{Energy $\Delta_2$ for a pair of holes as a function of hole 
separation, obtained by ED for clusters of 12, 18 and 24 sites with two 
impurities. }
\label{fig:8} 
\end{figure} 

\section{Hole repulsion} 
 
In this section we present results characterizing the behavior of an 
even--site cluster doped with two impurities. A comparison with the 
undisturbed system\cite{rlblps} reveals that the average energy per 
remaining site not only increases in the presence of impurities, but 
increases further as the impurities are brought closer together. Both 
of these features are evident from the ED results of Fig.~8, which shows 
the energies of all different two--hole states available for the relevant 
system sizes. The higher energy per site of doped systems confirms, by 
considering the contributing processes, that the net loss of resonance 
energy for dimers constrained by the presence of impurities to occupy 
only the bonds on which they are essentially ``frozen'' (Sec.~III) is not 
compensated by the gains in dimerization energy for those bonds which 
are enhanced. 

The effective hole repulsion is most unusual in low--dimensional spin 
systems,\cite{rbhsl,rdrs} and in fact we are unaware of analogous results 
for any model other than the frustrated chain.\cite{rrnm} 
This repulsion of holes constitutes the central result of our analysis:
particularly in the context of short--range RVB wave functions, it 
contradicts directly the most straightforward expectation for the 
behavior of doped holes, namely that when mobile these would choose to 
minimize the net disruption of the system by occupying the same dimer, 
and thus would exhibit a pairing attraction.\cite{ra} The consequences 
for mobile impurites with repulsive interactions may be expected to 
include hole crystallization, phase separation and the possibility of 
transitions between metallic and insulating phases, and would bear 
considerable further investigation.

\begin{figure}[t!] 
\psfig{figure=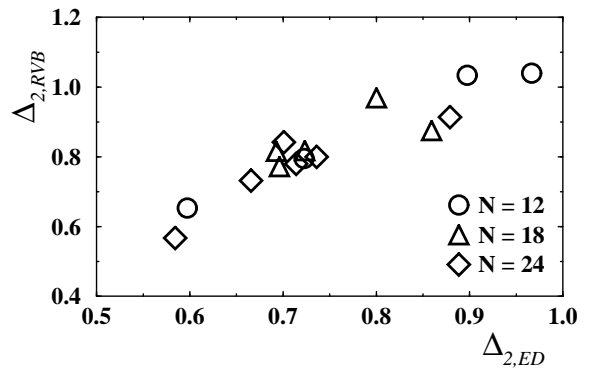,width=5.0cm,angle=270}
\medskip
\caption{Comparison of energies $\Delta_{2,{\rm ED}}$ and $\Delta_{2,{\rm 
RVB}}$ obtained respectively from ED and from calculations in the dimer--RVB 
basis for a pairs of holes of different separations on clusters of 12, 
18, and 24 sites with two impurities. }
\label{fig:9} 
\end{figure} 

\subsection{Dimer--RVB Basis}

We illustrate in Fig.~9 the efficacy of the dimer--RVB basis for the 
purposes of comparing ground--state energies. The correspondence between 
the dimer--RVB and ED results is shown by comparing the energies of all 
two--hole states in clusters of 12, 18 and 24 sites with two impurities, 
which represent all system sizes accessible by ED. Full details of the 
extraction of the energy values $\Delta_2$ in Fig.~9 are given below. 
Although an inspection of the abcissae reveals that the agreement between 
individual pairs of values is not as convincing as for the spin correlation 
functions (Sec.~III), the main difference between the two sets of data 
may be reduced to a relative upward shift of all the RVB energies. 
The direct linear correlation between ED and dimer--RVB results (alluded 
to in Sec.~II) demonstrates unambiguously that comparing the energies of 
different configurations of the two impurities, which is the primary goal 
of the energy analysis to follow, is a valid qualitative exercise when 
performed using the dimer--RVB results.

We show first the spin--spin correlation functions in the presence 
of a pair of impurities (Fig.~10). When considering the overlap of 
single--impurity dimerization patterns (Sec.~III), it is evident that 
these are not fully mutually compatible. In fact, one finds by inspection 
that not only the configuration illustrated in Fig.~10 but all possible 
configurations of two impurities on the kagome lattice are in some measure
incompatible. Comparison with other quantum spin systems suggests that 
this complete incompatibility is very specific to the kagome geometry, 
and its extent may be quantified by extracting the effective interaction 
energy of the holes (Fig.~11). In cluster calculations the frustration is 
also shown by the presence of relatively large local moments induced on 
the sites at the mismatch between dimerization regimes ({\it cf.}~Fig.~6). 

\begin{figure}[t!] 
\centerline{\psfig{figure=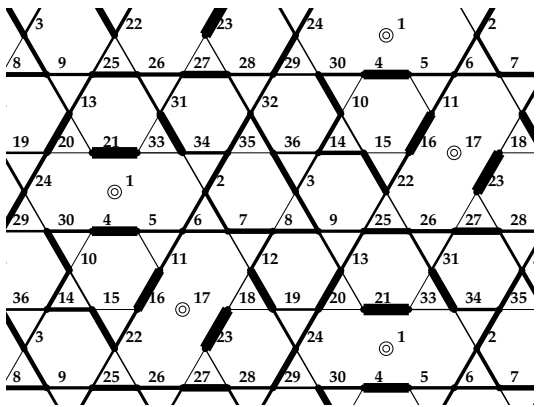,width=7.0cm,angle=0}}
\medskip
\caption{Spin--spin correlations for two impurities in a 36--site cluster, 
obtained in the dimer--RVB basis. Bar width represents the strength of 
correlation functions for each bond on a linear scale where the largest 
value is $- 0.66$. }
\label{fig:10} 
\end{figure} 
 
\subsection{Interaction between impurities}

Fig.~11 shows twice the energy per hole as a function of separation for all 
configurations of two impurities on 30-- and 36--site clusters, obtained 
in the dimer--RVB basis. We have analyzed the data in the following manner. 
$E_0$ is the ground--state energy of each cluster in the absence of 
impurities, and is computed for even clusters then extrapolated for 
odd clusters from the results for systems of sizes $N - 3$ and $N + 3$. 
$E_1$ is the ground--state energy in the presence of a single impurity, 
which is computed for odd clusters and may then be extrapolated for 
even clusters from the results for systems of sizes $N - 3$ and $N + 3$.
$E_2$ is the ground--state energy with two impurities, and again may be 
computed directly for even clusters and extrapolated for odd clusters.
We then define 
\begin{equation}
\Delta_1 = E_1 - E_0
\label{ed1}
\end{equation}
as the energy of one hole, 
\begin{equation}
\Delta_2 = E_2 - E_0
\label{ed2}
\end{equation}
as the energy of a two--hole state, and 
\begin{equation}
\epsilon = \Delta_2 - 2 \Delta_1 = E_2 - 2 E_1 + E_0
\label{ede}
\end{equation}
as the energy of interaction between two holes. Because of the extrapolation 
procedure it is important that the data sets for different system sizes be 
comparable. While certain problems arise for some system sizes due to the 
relatively large influence of the boundary conditions, we find the process 
to be valid for systems in excess of 27 sites. In this connection we note 
that the two--hole energies $\Delta_2$ shown in Fig.~11 are independent of 
the extrapolation between different system sizes, and it is only the direct 
comparison with $\Delta_1$ for the 33--site system which requires 
validation of the procedure.

\begin{figure}[t!] 
\psfig{figure=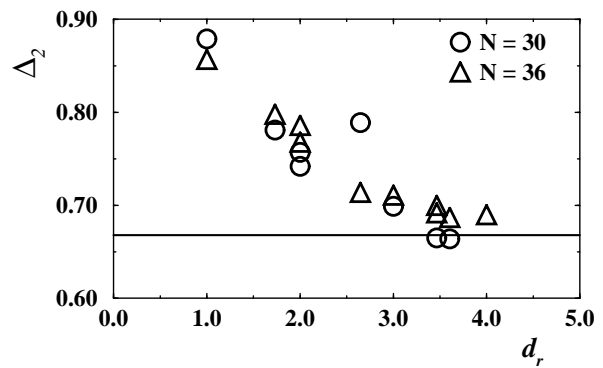,width=5.0cm,angle=270}
\medskip
\caption{Energies $\Delta_2$ for two--hole states as a function of hole
separation for all configurations of two impurities on 30-- and 36--site 
clusters, obtained by dimer--RVB calculations. The solid line marks twice 
the single--hole energy 2$\Delta_1$ relevant for comparison, taken from 
the 33--site cluster. }
\label{fig:11} 
\end{figure} 

It is clear that the effective mutual repulsion decays only slowly. Because 
of the restricted sizes of the acccessible clusters we are unable to 
speculate on the functional form of this decay, and in particular on the 
possibility that it may be long--ranged. We note only that for short 
distances (3--4 lattice constants) it appears to exhibit an approximate 
$1/d_r$ dependence, beyond which a steeper decrease is observed which may 
be either an intrinsic property of the interaction or a consequence of the 
system size. The form of the decay appears similar to that of the bond spin 
correlation functions towards their undisturbed value [Fig.~5(b)]. 

At the phenomenological level, a repulsive interaction between two 
impurities may be regarded as surprising in that their interference might 
have been expected to suppress the freezing of individual dimer patterns, 
leading to a restoration of resonance energy. Our findings to the contrary 
indicate that the dominant effect of increasing increasing impurity 
separation lies in the reduction in strength of the partially frozen 
single--impurity dimerization patterns at the boundaries where they are 
forced to mesh, which reduces the corresponding energetic penalty. Given 
the slow decay of the dimer--dimer correlations, the possibility remains 
that the net repulsive interaction between holes could be long--ranged. 
Such an unexpected result may indeed be connected with a true deconfinement 
of spinon--like excitations for the kagome system, which would confirm 
speculation to this effect in 2d.\cite{rsv} We comment further on the 
issue of deconfinement below. 

\section{Conclusion} 

We have illustrated a variety of novel phenomena associated with the 
effects of static, spinless impurities doped into the $S = 1/2$ 
antiferromagnetic Heisenberg model in kagome geometry. By analysis of 
spin--spin correlation functions and of the local magnetization patterns 
induced around impurities, we find that, despite the very short spin--spin 
correlation lengths of the highly frustrated system, dimer--dimer 
correlations develop over considerable distances due to a hole--induced 
freezing of their resonance. Furthermore, as a consequence of the 
continuum of states in the low--lying singlet manifold, there is no 
evidence for free local moments in the vicinity of impurity sites. 

Most importantly, we have shown from the total energies of doped systems 
that impurities in the kagome lattice have an effective repulsive 
interaction, a phenomenon unique to our knowledge and in direct 
contradiction to the conventional expectation for short--range RVB 
systems of a net pairing attraction. Furthermore, this effective 
repulsion is found to be rather long--ranged in nature, a most unusual 
result which is presumably related with the very strong frustration 
inherent to the kagome geometry. 

Our comparisons of doping effects provide additional justification for the 
validity of the short--range dimer--RVB framework as a suitable description 
of the ground--state singlet manifold in the kagome antiferromagnet. We 
have shown that the dimer--RVB basis provides a valuable means of probing 
the nature of spin and charge degrees of freedom in a further class of 
quantum magnets. This realization of a highly resonant RVB state, arising 
as a consequence of the extremely frustrated nature of the system, also 
presents an important contrast to other low--dimensional spin systems 
which are known to be described by RVB--type wave functions, particularly 
with regard to the localization of free spins and to the pairing of holes.

We summarize only briefly the consequences of our analysis for experiment. 
The majority of known kagome materials do not have complete filling of the 
lattice sites by a single value of the spin $S$, and as such the study of 
impurities is a significant component of their characterization. Nonmagnetic 
impurities in the kagome lattice have the important property that they do 
not generate localized spins at the vacant site, and do generate by the 
nature of their extended range of influence a significant number of 
satellites for each peak. On heuristic grounds we believe that our analysis 
of the qualitative features to be expected in the $S = 1/2$ system are 
relevant also for $S = 3/2$, and perhaps for higher half--integral spins. 

Finally, on the issue of deconfinement in highly frustrated systems of 
dimensionality higher than one, we have shown clearly from the presence 
of the mutual repulsive interaction between holes that charge degrees 
of freedom are deconfined. Further, the absence of localized $S = 1/2$ 
degrees of freedom around vacancies in the lattice demonstrates that 
spin and charge are deconfined. However, on the question of whether 
the spin excitations of the undoped system are appropriately described 
by deconfined spinon excitations, our results do not provide a proof of 
this hypothesis. 

\acknowledgments 
 
We are grateful to P. Sindzingre for helpful provision of ED spectra for 
undoped clusters. This work was supported by the Swiss National Science 
Foundation and by the French Centre National de Recherche Scientifique.


\begin{thebibliography}{99}
\bibitem{footnote1} The behavior of the number of inequivant sites as a 
function of the spin sector arises from the presence of an additional 
permutation symmetry among the sites for the 27--site cluster with the 
chosen boundary conditions. This symmetry does not in general commute 
with the two reflection symmetries of the cluster, and so cannot be 
obtained simultaneously with these. However, in the majority of spin 
sectors it is possible to obtain all three symmetries simulataneously, 
making sites $i = 2$ and 3, $i = 5$ and $5^{\prime}$, and $i = 4$ and 7 
equivalent in these cases.
\bibitem{footnote2} A further heuristic justification for the relevance 
of the current results at least to the $S = 3/2$ case may be obtained 
by considering that the kagome geometry supports a non--degenerate 
valence--bond--solid state\cite{raklt} for spins $S = 2$. In an RVB basis 
a system of spins $S = 3/2$ is equivalent to a set of dimer vacancies in 
this background which have exactly the same degeneracy as the dimers in 
the $S = 1/2$ system.
\bibitem{rrnm} Reexamination of results obtained by two of the 
authors\protect{\cite{rnm}} reveals that a mutual repulsion of 
impurities also occurs in the spin chain with frustrating next--neighbor 
interactions.
\end{thebibliography}
\end{document}